\documentclass[prl,aps,superscriptaddress,twocolumn,bibnotes,footinbib]{revtex4-2}
\usepackage{amsmath,amssymb}
\usepackage[colorlinks,bookmarks=false,citecolor=magenta,linkcolor=magenta,urlcolor=magenta]{hyperref}
\usepackage{tikz}
\usepackage{mathtools}
\usepackage{bm}
\usepackage{ragged2e}
\usepackage{graphicx}

\newcommand{\myroman}[1]{\uppercase\expandafter{\romannumeral#1}}
\usepackage[normalem]{ulem}

\begin{document}

	\title{Observation of multiple steady states with engineered dissipation}

	\author{Li Li}
	\thanks{These authors contributed equally to this work.}
	\affiliation{Institute of Physics, Chinese Academy of Sciences, Beijing 100190, China}
	\affiliation{School of Physical Sciences, University of Chinese Academy of Sciences, Beijing 100190, China}

	\author{Tong Liu}
	\thanks{These authors contributed equally to this work.}
	\affiliation{Institute of Physics, Chinese Academy of Sciences, Beijing 100190, China}
	
	\author{Xue-Yi Guo}
	\affiliation{Institute of Physics, Chinese Academy of Sciences, Beijing 100190, China}

	\author{He Zhang}
	\affiliation{Institute of Physics, Chinese Academy of Sciences, Beijing 100190, China}
	\affiliation{School of Physical Sciences, University of Chinese Academy of Sciences, Beijing 100190, China}

	\author{Silu Zhao}
	\affiliation{Institute of Physics, Chinese Academy of Sciences, Beijing 100190, China}
	\affiliation{School of Physical Sciences, University of Chinese Academy of Sciences, Beijing 100190, China}

	\author{Zhongcheng Xiang}
	\email{zcxiang@iphy.ac.cn}
	\affiliation{Institute of Physics, Chinese Academy of Sciences, Beijing 100190, China}
	\affiliation{School of Physical Sciences, University of Chinese Academy of Sciences, Beijing 100190, China}
	\affiliation{Beijing Academy of Quantum Information Sciences, Beijing 100193, China}
	\affiliation{Hefei National Laboratory, Hefei 230088, China}
	\affiliation{CAS Center of Excellence for Topological Quantum Computation, University of Chinese Academy of Sciences, Beijing 100190, China}
	\affiliation{Songshan Lake Materials Laboratory, Dongguan 523808, Guangdong, China}

	\author{Xiaohui Song}
	\affiliation{Institute of Physics, Chinese Academy of Sciences, Beijing 100190, China}
	\affiliation{Beijing Academy of Quantum Information Sciences, Beijing 100193, China}
	\affiliation{Hefei National Laboratory, Hefei 230088, China}
	
	\author{Yu-Xiang Zhang}
	\affiliation{Institute of Physics, Chinese Academy of Sciences, Beijing 100190, China}
	\affiliation{School of Physical Sciences, University of Chinese Academy of Sciences, Beijing 100190, China}
	\affiliation{Hefei National Laboratory, Hefei 230088, China}

	\author{Kai Xu}
	\email{kaixu@iphy.ac.cn}
	\affiliation{Institute of Physics, Chinese Academy of Sciences, Beijing 100190, China}
	\affiliation{School of Physical Sciences, University of Chinese Academy of Sciences, Beijing 100190, China}
	\affiliation{Beijing Academy of Quantum Information Sciences, Beijing 100193, China}
	\affiliation{Hefei National Laboratory, Hefei 230088, China}
	\affiliation{CAS Center of Excellence for Topological Quantum Computation, University of Chinese Academy of Sciences, Beijing 100190, China}
	\affiliation{Songshan Lake  Materials Laboratory, Dongguan 523808, Guangdong, China}

	\author{Heng Fan}
	\email{hfan@iphy.ac.cn}
	\affiliation{Institute of Physics, Chinese Academy of Sciences, Beijing 100190, China}
	\affiliation{School of Physical Sciences, University of Chinese Academy of Sciences, Beijing 100190, China}
	\affiliation{Beijing Academy of Quantum Information Sciences, Beijing 100193, China}
	\affiliation{Hefei National Laboratory, Hefei 230088, China}
	\affiliation{CAS Center of Excellence for Topological Quantum Computation, University of Chinese Academy of Sciences, Beijing 100190, China}
	\affiliation{Songshan Lake Materials Laboratory, Dongguan 523808, Guangdong, China}

	\author{Dongning Zheng}
	\affiliation{Institute of Physics, Chinese Academy of Sciences, Beijing 100190, China}
	\affiliation{School of Physical Sciences, University of Chinese Academy of Sciences, Beijing 100190, China}
	\affiliation{Hefei National Laboratory, Hefei 230088, China}
	\affiliation{CAS Center of Excellence for Topological Quantum Computation, University of Chinese Academy of Sciences, Beijing 100190, China}
	\affiliation{Songshan Lake  Materials Laboratory, Dongguan 523808, Guangdong, China}

	\begin{abstract}
	
	Simulating the dynamics of open quantum systems is essential in achieving practical quantum computation and understanding novel nonequilibrium behaviors.
	However, quantum simulation of a many-body system coupled to an engineered reservoir has yet to be fully explored in present-day experiment platforms.
	In this work, we introduce engineered noise into a one-dimensional ten-qubit superconducting quantum processor to emulate a generic many-body open quantum system.
	Our approach originates from the stochastic unravellings of the master equation.
	By measuring the end-to-end correlation, we identify multiple steady states stemmed from a strong symmetry, which is established on the modified Hamiltonian via Floquet engineering.
	Furthermore, we find that the information saved in the initial state maintains in the steady state driven by the continuous dissipation on a five-qubit chain.
	Our work provides a manageable and hardware-efficient strategy for the open-system quantum simulation.

	\end{abstract}

	\maketitle

	\emph{Introduction.---}The interactions between the quantum system and its environment lead to the decoherence, which disrupts the many-body coherence lying at the heart of quantum information processing (QIP)~\cite{Nielsen2012}.
    Despite the detrimental effects of uncontrolled dissipation, engineering the system-environment coupling reshapes the dissipation into a powerful tool in various fields of QIP, including preparation and stabilization of entangled states~\cite{Diehl2008,PhysRevA.78.042307,Verstraete2009,Lin2013,PhysRevLett.115.240501,PhysRevX.6.011022,Ma2019}, active reset of qubits~\cite{Valenzuela2006,PhysRevLett.110.120501,PhysRevLett.121.060502}, and quantum error-correction~\cite{Leghtas2015,PhysRevLett.116.150501,PhysRevX.8.021005,Gertler2021}.
	The dynamic interplay between the system Hamiltonian and dissipation gives rise to exotic nonequilibrium scenarios such as dissipation phase transition~\cite{PhysRevLett.101.105701,PhysRevA.86.012116,PhysRevA.95.012128,PhysRevLett.122.110405,PhysRevLett.123.173601} and dissipative time crystals~\cite{PhysRevLett.120.040404,PhysRevLett.122.015701,PhysRevLett.129.250401,PhysRevLett.130.130401}.
	Recent advances in quantum simulators offer an opportunity to investigate these intriguing nonequilibrium phenomena experimentally~\cite{PhysRevX.7.011012,PhysRevX.7.011016,Fink2017,PhysRevLett.127.043602,Taheri2022}.

\begin{figure}[!h]
		\centering
		\includegraphics[width=0.9\linewidth]{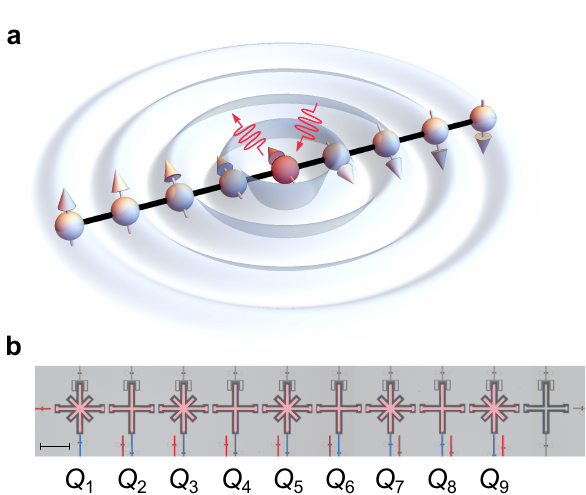}
		\caption{\quad Spin chain with a local dissipation and device. (a) An array of nine qubits with nearest-neighbor couplings. The dissipation on the center qubit impels the system to steady states with long-range coherence. (b) Optical picture of the ten-qubit superconducting quantum processor with highlighting circuit elements. Qubits (pink) are labeled from $Q_1$ to $Q_9$ and can be controlled by respective XY lines (red) and Z lines (blue). Scale bar in the lower left corner, 0.2 mm.}
		\label{fig:sample}
	\end{figure}

	As a paradigmatic configuration, a spin-1/2 chain integrated with local dissipations at edges allowing exact analyses has been utilized to explore nonequilibrium heat and spin transport~\cite{Prosen2008,nidari2010,PhysRevLett.106.217206,PhysRevLett.106.220601,RevModPhys.94.045006}. 
	Recent studies reveal that multiple steady states emerge for pump and loss at center spins instead of boundaries, which is rooted in the strong symmetry of setup depicted in Fig.~\ref{fig:sample}a~\cite{PhysRevLett.125.240404,PhysRevResearch.3.L012016}.
	Such kind of strong symmetry in open quantum systems has been unveiled in theoretical works~\cite{Bua2012,PhysRevB.90.125138}, but is lack of experimental realizations. 

	Here, we report our experiment in probing multiple steady states induced by a strong symmetry on a one-dimensional superconducting quantum processor with nine qubits as shown in Fig.~\ref{fig:sample}b.
	The qubits used in the experiment are labeled by $Q_i$ with $i\in\{1,2,\dots,9\}$.
	Each qubit can be addressed by individual control lines.
	Based on the interpretation of open quantum dynamics in terms of stochastic wave functions,
	we engineer a stochastic Hamiltonian to mimic the evolution featuring controllable dissipation by averaging over a set of unitary evolutions.
	Our protocol can be efficiently extended to multiple dissipations on the current noisy intermediate-scale quantum devices, without the need for ancillary qubits~\cite{Barreiro2011,PhysRevLett.127.020504}, intense classical resources~\cite{PhysRevLett.125.010501,Haug2022}, or dissipative elements~\cite{Ma2019}.
	Incorporating the Floquet engineering to modify the interaction strengths of a nine-qubit chain, we show that the system evolves to different steady states specified by the end-to-end correlation, initialized in distinct symmetry sectors.
	Our results demonstrate that superconducting circuit is a promising platform to examine nontrivial properties of the nonequilibrium many-body system benefiting from its high flexibility and manipulability.

	\begin{figure}[!ht]
		\centering
		\includegraphics[width=0.9\linewidth]{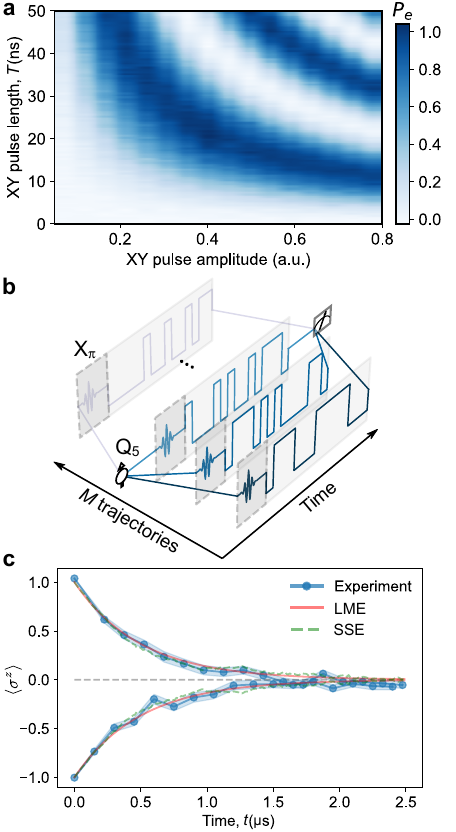}
		\caption{\quad Simulation of the Lindblad master equation Eq.~(\ref{eq:thermal}).
				(a) The Rabi oscillation for calibration of the XY pulse amplitude. 
				(b) A Schematic of XY pulse sequences for $M$ trajectories. 
				(c) The evolutions of $\langle\sigma^z \rangle$ in experiment after 100 repetitions,
					governed by Lindblad master equation, and simulated by stochastic Schr\"odinger equation.
					The shaded light blue region represents the standard error of the mean over trajectories in the experiment.}
		\label{fig:single_qubit}
	\end{figure}
	\emph{Results.---}We start with the simulation of the following Lindblad master equation (LME) of a single qubit~\cite{Lindblad1976,Daley2014}, 
	\begin{align}
		\frac{d\rho}{dt} & = \sum_{j=1,2} L_j\rho L_j^\dagger - \frac{1}{2}\{L_j^\dagger L_j, \rho\},
		\label{eq:sq_eq}
	\end{align}
	where $\rho$ is the density matrix of the qubit, $L_1 = \sqrt{\gamma_+}\sigma^+ = \sqrt{\gamma_+}|e\rangle\langle g|$ and $L_2 = \sqrt{\gamma_-}\sigma^- = \sqrt{\gamma_-}|g\rangle\langle e|$ are pump and loss operators, respectively, with $\gamma_+$ ($\gamma_-$) being the pump (loss) rate, and $|g\rangle$ ($|e\rangle$) is the ground (excited) state of qubit.
	Here, we set $\gamma_+ = \gamma_- = \gamma$. A more general case $(\gamma_- > \gamma_+)$ is discussed in the Supplementary Materials~\cite{sm}.
	Now Eq.~(\ref{eq:sq_eq}) can be rewritten as
	\begin{equation}
		\dot\rho = \frac{\gamma}{2}\left(\sigma^x\rho\sigma^x + \sigma^y\rho\sigma^y - 2\rho\right).\label{eq:thermal}
	\end{equation}
	The form of Eq.~(\ref{eq:thermal}) reminds us of that $\rho$ is the ensemble average of stochastic wave function $|\psi\rangle$, i.e., $\rho = \overline{|\psi\rangle\langle\psi|}$,
	where the overline denotes the average over stochastic realizations~\cite{PhysRevA.63.012106,PhysRevLett.118.140403,PhysRevLett.122.050501}.
	The stochastic wave function $|\psi\rangle$ is governed by the following stochastic Schr\"odinger equation (SSE)
	\begin{align}
		\frac{d}{dt}|\psi\rangle &= -iH_S(t)|\psi\rangle \notag \\
				 & = -i\left[\sqrt{\frac{\gamma}{2}}\xi_1(t)\sigma^x + \sqrt{\frac{\gamma}{2}}\xi_2(t)\sigma^y\right]|\psi\rangle ,
				 \label{eq:sse}
	\end{align}
	where $\xi_1(t)$ and $\xi_2(t)$ are two independent real Gaussian processes satisfying $\overline{\xi_\alpha(t)\xi_\beta(t')} = \delta_{\alpha\beta}\delta(t-t')$ and $\overline{\xi_\alpha} = 0$ for $\alpha, \beta = 1,2$.

	\begin{figure*}[!ht]
		\centering
		\includegraphics[width=0.95\textwidth]{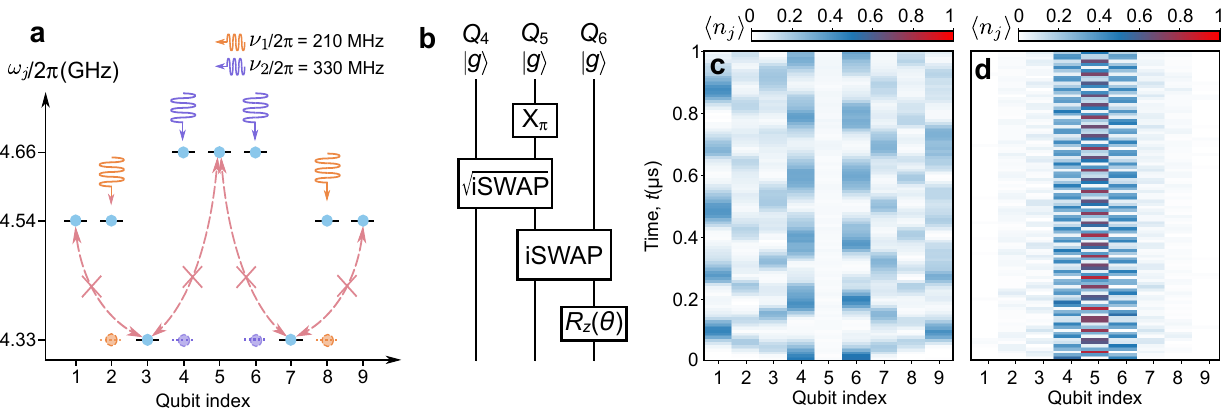}
		\caption{\quad Floquet engineering and the evolution of a pair of Bell state in a 1D array with nine qubits. (a) Idle frequencies and flux modulations of nine qubits (b) The circuit to prepare a pair of Bell state between $Q_4$ and $Q_6$. (c)(d) The time evolution of density distribution $\langle n_j\rangle$ under the periodic driving for Bell phase $\varphi=\pi$ or $\varphi=0$.}
		\label{fig:quantum_walk}
	\end{figure*}

	Although $H_S(t)$ is a Hermitian Hamiltonian, the faithful generation of ideal Gaussian processes in Eq.~(\ref{eq:sse}) is infeasible in experiments due to the finite bandwidth of arbitrary wave generators. 
	Inspired by the numerical techniques of differential equations, we adopt the Euler's method to simulate Eq.~(\ref{eq:sse}) by slicing each trajectory into $N$ sections divided by time intervals of a small duration $\Delta t$.
	The evolution of the wave function in the $i$th section $|\psi_i(t)\rangle$ is given by
	\begin{equation}
		\frac{d}{dt}|\psi_{i}(t)\rangle = -i\left[\sqrt{\frac{\gamma}{2\Delta t}}\eta_1^i\sigma^x + \sqrt{\frac{\gamma}{2\Delta t}}\eta_2^i\sigma^y\right]|\psi_i(t)\rangle,\label{eq:sse_discrete}
	\end{equation}
	with the initial condition $|\psi_i(0)\rangle = |\psi_{i-1}(\Delta t)\rangle$, where $\eta_1^i$ and $\eta_2^i$ are two random variables following a discrete distribution $P(\eta_{1(2)} = 1) = P(\eta_{1(2)} = -1) = 1/2$~\cite{sm}. 
	By sampling $2N$ variables $\{\eta_1^1,\dots,\eta_1^N,\eta_2^1,\dots,\eta_2^N\}$, we can determine a trajectory and generate the corresponding XY pulse sequences according to Eq.~(\ref{eq:sse_discrete}).
	Note that the amplitude of XY pulse keeps fixed while the phase in each section is uniformly chosen from $\{\pi/4, 3\pi/4, 5\pi/4, 7\pi/4\}$.
	The mapping between the amplitude of XY pulse and $\gamma$ can be calibrated in prior via the Rabi oscillation, where we apply a rectangular XY pulse to the qubit initialized as $|g\rangle$, and measure the excitation probability $P_e$ versus the XY pulse length $T$, as shown in Fig~\ref{fig:single_qubit}a.
	After sampling $M$ trajectories, the dynamics of an observable $O$ can be estimated by $\sum_{j=1}^M \langle \psi^{(j)}|O|\psi^{(j)}\rangle/M$ where $|\psi^{(j)}\rangle$ is the $j$th trajectory.

	We testify our scheme on the qubit $Q_5$ by measuring the evolution of $\sigma^z\equiv |e\rangle\langle e| - |g\rangle\langle g|$ from two initial states $|g\rangle$ and $|e\rangle$ with $\Delta t = 7.5$ ns, $\gamma = 1$ MHz and $M=100$, as illustrated in Fig.~\ref{fig:single_qubit}b.
	The results are presented in Fig.~\ref{fig:single_qubit}c and compared with numerical results calculated by LME and SSE.
	We find that the experiment results are in good agreements with simulations in a duration of 2.5 $\mu$s, and converge to a steady state $\rho_\mathrm{ss}$ in which $\langle \sigma^z \rangle=0$ irrespective of the initial states chosen.

	Whereas the dissipation drives the qubit into a thermal state represented by $(\gamma_+|e\rangle\langle e| + \gamma_-|g\rangle\langle g|)/(\gamma_+ + \gamma_-)$, it is surprising that the introduction of local dissipations into the center spin of a XX chain consisting of an odd number of spins results in multiple long-range coherent steady states~\cite{PhysRevLett.125.240404,PhysRevResearch.3.L012016}. 
	The Hamiltonian $H$ of a XX chain exhibiting a reflection symmetry reads ($\hbar = 1$)
	\begin{equation}
		H = \sum_{i=1}^{L-1}J_{i,i+1}(\sigma_i^+\sigma_{i+1}^- + \sigma_i^-\sigma_{i+1}^+),\label{eq:Hamiltonian}
	\end{equation}
	where $L$ is the length of the chain and $J_{i,i+1}/2\pi\approx 11$ MHz is the nearest-neighbor (NN) interaction strength with $J_{i,i+1}=J_{L-i,L+1-i}$.
	A hidden symmetry arising from a operator $C^2$ restricts the dynamics into different symmetry sectors where
	\begin{equation}
		C = -\frac{1}{2} + \sum_{k=1}^l f_k^\dagger f_{L+1-k}, 
	\end{equation}
	$f_k$ is the fermionic operator at site $k$ derived from Jordan-Wigner transformation~\cite{Jordan1928} and $l=(L-1)/2$.
	However, the symmetry is broken when the Hamiltonian contains next-nearest-neighbor (NNN) interactions of which strengths $J_{i,i+2}/2\pi$ amount to 1 MHz in our processor.
	We suppress the unwanted NNN interactions via Floquet engineering as illustrated in Fig.~\ref{fig:quantum_walk}a~\cite{PhysRevApplied.10.054009,PhysRevLett.123.080501,PhysRevLett.129.160602}.
	The idle frequencies $\omega_j/2\pi$ of nine qubits are aligned to three specific frequencies 4.33 GHz, 4.54 GHz and 4.66 GHz, which turns off the NNN interactions (red dashed lines in Fig.~\ref{fig:quantum_walk}a) between the following pairs of qubits: $Q_1$ and $Q_3$, $Q_3$ and $Q_5$, $Q_5$ and $Q_7$, and $Q_7$ and $Q_9$.
	Then we apply ac magnetic fluxes to modulate the frequencies of $Q_2$, $Q_4$, $Q_6$, and $Q_8$ as $\tilde \omega_j(t) = \omega_j + \varepsilon_j\sin(\nu_j t + \phi_j)$ for $j = 2, 4, 6$ and 8,
	with $\varepsilon_j$, $\nu_j$ and $\phi_j$ being the modulation amplitude, frequency, and phase, respectively.
	When the modulation frequency is far larger than the NN interaction strength, the rapid oscillation induces a set of sidebands $\omega_j + m\nu_j$, where $m$ is an integer.
	To initiate the interactions between adjacent qubits operating at distinct idle frequencies, the modulation frequencies $\nu_j$ are equal to the frequency detuning $\Delta/2\pi = |\omega_j - \omega_{j+1}|/2\pi = 210$ MHz or $|\omega_j - \omega_{j-1}|/2\pi = 330$ MHz.
	The modulation amplitudes $\varepsilon_j$ are adjusted to rectify the minor coupling disorder in the processor, ensuring that the NN interaction strengths remain primarily symmetric.
	The NNN interaction between the qubits $Q_2$ ($Q_6$) and $Q_4$ ($Q_8$) is also undermined by the Floquet engineering, and
	the remaining NNN interaction between $Q_4$ and $Q_6$ retains the symmetry as both qubits are equidistant from $Q_5$~\cite{PhysRevLett.125.240404}.

	To observe the different steady states, the initial state must be prepared within the diverse eigenspaces of $C^2$.
	The eigenstate of $C$ is given by~\cite{PhysRevLett.125.240404}
	\begin{equation}
		|\{v_{k,\pm}, n_0\}\rangle = (f_0^\dagger)^{n_0}\prod_{k=1}^l\prod_{s=\pm}(a_{k,s}^\dagger)^{\nu_{k,s}}|0\rangle,
	\end{equation}
	with eigenvalue $\lambda = \sum_{k=1}(\nu_{k,+} - \nu_{k,-}) + n_0 - 1/2$ where $|0\rangle$ is the vacuum state, $n_0$ and $\nu_{k,\pm} \in \{0, 1\}$, and $a_{k,\pm} = (f_{k}\pm f_{L+1-k})/\sqrt 2$ for $k = 1, \dots, l$.
	$C^2$ shares the same eigenstates with $C$ but possesses $(l+1)$ distinguishable eigenvalues $(\eta + 1/2)^2$ with $\eta = 1, \dots, l$.
	Considering the degeneracy of eigenvalues, it is possible to traverse all eigenspaces of $C^2$ by increasing $v_{k,-}$ from 0 to $l$ with $n_0=0$ and $v_{k,+}=0$ for $k =1 ,\dots, l$. 
	Hence, by defining Bell state creating operators between qubits $Q_{l-k + 1}$ and $Q_{L-l+k}$, denoted as $b_{k,\pm}^\dagger = (\sigma_{l-k+1}^+ \pm \sigma_{L-l+k}^+)/\sqrt 2$ for $k = 1,\dots,l$, we can generate $(l+1)$ states $\{|\phi_\eta\rangle\}$ belonging to distinct symmetry sectors
	\begin{equation}
		|\phi_0\rangle \equiv \bigotimes_{j=1}^L|g_j\rangle,\, |\phi_\eta\rangle\equiv \prod_{k=1}^\eta b^\dagger_{k, (-)^k}|\phi_0\rangle, \, \eta = 1,\dots, l,
	\end{equation}
	where $|g_j\rangle$ is the ground state of $Q_j$.
	Exploiting the circuit shown in Fig.~\ref{fig:quantum_walk}b, we implement $|\phi_1\rangle$ with state fidelity more than $99.9\%$ characterized by the quantum state tomography in our processor~\cite{PhysRevA.64.052312}.
	The $R_z(\theta)$ gate in the circuit is defined as $R_z(\theta) \equiv \exp(-i\sigma^z\theta/2)$ to tune the phase of Bell state $|\Psi(\varphi)\rangle\equiv \frac{1}{\sqrt 2}(|e_{l-1},g_{l+1}\rangle + e^{i\varphi}|g_{l-1},e_{l+1}\rangle)\bigotimes_{j\ne l-1,l+1 }^{L}|g_j\rangle$ as $\varphi = 0$ or $\varphi = \pi$, corresponding to symmetry sectors $\eta = 0$ or $\eta=1$, respectively.
	We also implement $|\phi_2\rangle$ and $|\phi_3\rangle$ by repeating generating Bell states  with staggered phases through the same circuit and iSWAP gates~\cite{sm}.

	Figure~\ref{fig:quantum_walk}c shows the single excitation density distribution $\langle n_j\rangle\equiv(1+\langle\sigma_j^z\rangle)/2$ launched from $|\phi_1\rangle$ under the periodic driving, where two excitations propagate towards two opposite directions with the same velocity due to the reflection symmetry, and swing between the boundary qubit and the center qubit $Q_5$.
	The occupation number of the center qubit is always nearly zero because of the destructive interference of two excitations, which can also be understood by the fact that $n_0=1$ is excluded from the eigenspace of symmetry sector $\eta=1$.
	On the contrary, when the phase of initial Bell state is zero, two excitations are almost confined between $Q_4$ and $Q_6$ in Fig.~\ref{fig:quantum_walk}d, which arises from the non-uniform effective interaction strengths between NN qubits.
	This phenomenon additionally facilitates the calibration of both the Bell state phase and the modulation phase of the longitudinal field in the experiment~\cite{sm}.

	\begin{figure}[!t]
		\centering
		\includegraphics[width=0.95\linewidth]{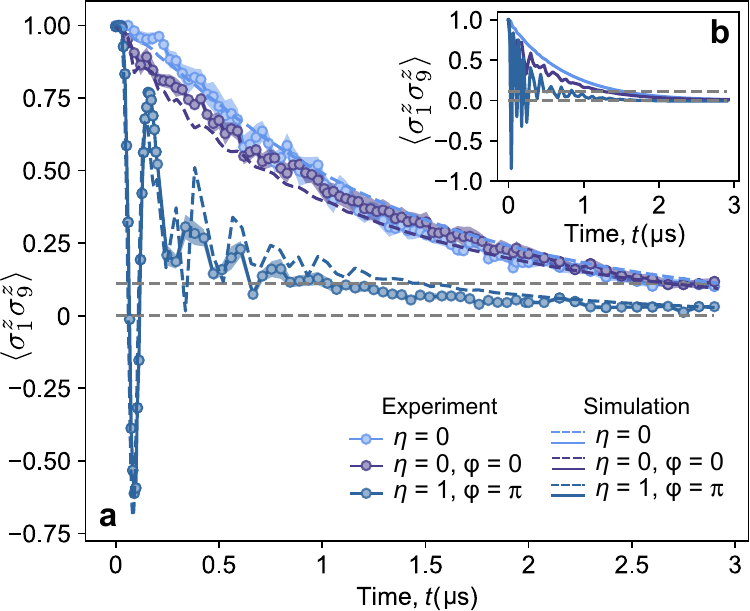}
		\caption{\quad The evolution of end-to-end correlation $\langle \sigma_1^z\sigma_L^z\rangle$ on a nine-qubit chain.
		The shaded regions surrounding the experiment data represent the standard error of the mean over trajectories in the experiment.
		(a) Three solid lines with circle markers correspond to the initial state $|\phi_0\rangle$, $|\Psi(0)\rangle$, and $|\Psi(\pi)\rangle$, respectively. Dashed lines are numerical simulations with $T_1=30$ $\mu$s and $T_\phi=20$ $\mu$s. 
		(b) Numerical simulations using the original device parameters without Floquet engineering.}
		\label{fig:9qubit_noisy}
	\end{figure}

	\begin{figure}[!t]
		\includegraphics[width=0.95\linewidth]{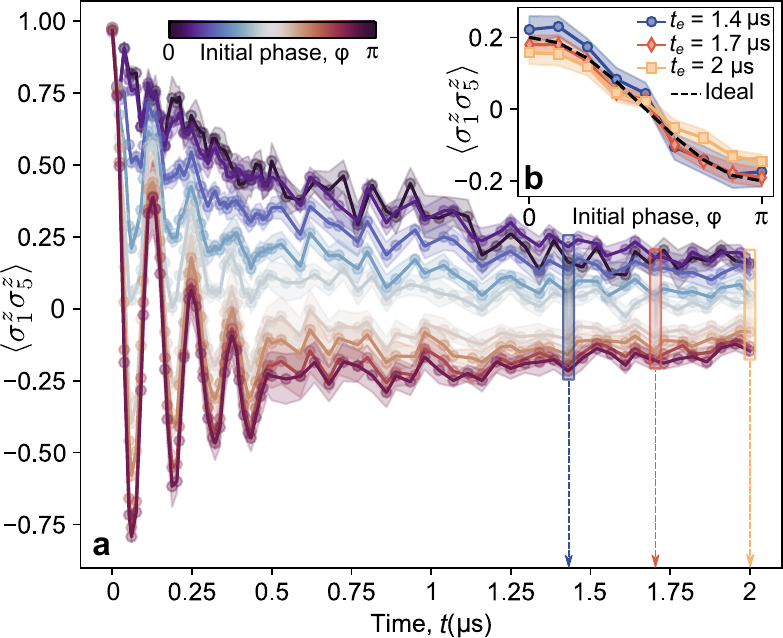}
		\caption{\quad The evolution of end-to-end correlation $\langle \sigma_1^z\sigma_L^z\rangle$ for Bell states with different phases on a five-qubit chain. 
		The shaded regions surrounding the experiment data represent the standard error of the mean over trajectories in the experiment.
		(a) The solid lines with circle markers correspond to the initial phases from $0$ to $\pi$ in increments $\pi/8$, respectively. (b) The values of $\langle\sigma_1^z\sigma_5^z\rangle$ evaluated at $t_e=1.4$ $\mu$s, $1.7$ $\mu$s and $2$ $\mu$s.}
		\label{fig:5qubit_noisy}
	\end{figure}

	Now we consider the dynamics of system when turning on the dissipation of $Q_5$.
	We measure the evolution of the end-to-end correlation $\langle\sigma_1^z\sigma_L^z \rangle$
	from three initial states by performing joint readouts of the qubits located at the ends of the nine-qubit chain as shown in Fig.~\ref{fig:9qubit_noisy}a, where we sample 10 trajectories for each evolution.
	For the initial states $|\phi_0\rangle$ and $|\Psi(0)\rangle$, which hold different numbers of excitation but belong to the same sector $\eta=0$, $\langle\sigma_1^z\sigma_L^z \rangle$ tends to the steady value $\langle\sigma_1^z\sigma_L^z\rangle_\mathrm{st}^0 = 1/L$, or $1/9$ for $L=9$~\cite{PhysRevLett.125.240404} indicated by the upper gray dashed line in the Fig.~\ref{fig:9qubit_noisy}a.
	For the other initial state $|\Psi(\pi)\rangle$ in the sector $\eta=1$, $\langle \sigma_1^z\sigma_N^z \rangle$ tends to the steady value $\langle\sigma_1^z\sigma_L^z\rangle_\mathrm{st}^1=(l-4)/Ll$, or $0$ for $L=9$~\cite{PhysRevLett.125.240404} indicated by the lower gray dashed line.
	The experimental results are consistent with the numerical simulations involving the energy relaxation time $T_1=30$ $\mu$s and Ramsey dephasing time $T_\phi=20$ $\mu$s for each qubit.
	In Fig~\ref{fig:9qubit_noisy}b, we show the numerical results simulated with the Hamiltonian built from the original device parameters without Floquet engineering.
	All three lines rapidly converges to zero owing to the vanishing of the symmetry caused by the NNN interaction and asymmetric NN interaction strengths.
	
	Finally, we explore the structure of degenerate steady states using 5 qubits $\{Q_3, Q_4, Q_5, Q_6, Q_7\}$ with the other qubits being far off-resonant.
	The steady value $\langle\sigma_1^z\sigma_L^z\rangle_\mathrm{st}(\varphi)$ corresponding to the initial state $|\Psi(\varphi)\rangle$ on a five-qubit chain is expected as $\cos\varphi/5$, derived from the combination of $\langle\sigma_1^z\sigma_L^z\rangle_\mathrm{st}^0$ and $\langle\sigma_1^z\sigma_L^z\rangle_\mathrm{st}^1$~\cite{sm}.
	The experimental results are shown in Fig.~\ref{fig:5qubit_noisy}a where we increase the initial phase $\varphi$ from 0 to $\pi$ in increments $\pi/8$ by the $R_z(\theta)$ gate.
	The evolution from the state $|\Psi(\varphi)\rangle$ tends to a steady value between two extreme steady values $\langle\sigma_1^z\sigma_L^z\rangle_\mathrm{st}^0$ and $\langle\sigma_1^z\sigma_L^z\rangle_\mathrm{st}^1$.
	In Fig.~\ref{fig:5qubit_noisy}b, we collect the values of $\langle \sigma_1^z\sigma_L^z\rangle$ evaluated at $t_e=1.4$ $\mu$s, $1.7$ $\mu$s, and $2$ $\mu$s for different initial phases.
	The data collected at $t_e=1.7$ $\mu$s is closest to the ideal result.
	Due to the accumulated decoherence errors, the end-to-end correlations at $t_e=2$ $\mu$s are smaller than those at $t_e=1.7$ $\mu$s, but the feature of the cosine function remains.
	These observations demonstrate that the phase information stored in the initial state can be preserved through the engineered dissipation, and $|\Psi(0)\rangle$ and $|\Psi(\pi)\rangle$ constitute a pointer basis for a classical bit~\cite{RevModPhys.75.715,PhysRevA.82.062306,PhysRevA.89.022118,PhysRevLett.116.240404}.

	\emph{Conclusion and discussion.---}We use a discretized SSE to simulate a class of LME by the associated stochastic Hamiltonian, and examine the protocol on a transmon qubit. 
	To observe the multiple steady states in our processor, we harness Floquet engineering to suppress undesirable NNN interactions and observe the quantum walk of a Bell state in a superconducting qubit chain.
	By tuning the phase of the Bell state and activating the dissipation, the end-to-end correlation of a nine-qubit chain converges to the steady value in the symmetry sector $\eta=0$ or $\eta=1$.
	We also show that the phase information in the initial state can be extracted from the steady state in a five-qubit array.
	Our method can also be extended to explore symmetry sectors with $\eta > 1$ by generating multiple pairs of Bell states and further increasing the ratio $U/J_{i,i+1}$~\cite{sm,PhysRevLett.129.160602}.
	In addition, our approach could have potential applications in probing the spreading of correlation in open quantum systems~\cite{PhysRevB.103.L020302} and diagnosing non-Markovian dynamics~\cite{RevModPhys.88.021002}.

	\section{Acknowledgments}
	Devices were made at the Nanofabrication Facilities at Institute of Physics, CAS in Beijing. This work was supported by: the National Natural Science Foundation of China (Grants No. 11875220, 11904393, 11934018, 12005155, 12047502, 12204528, 92065114, 92265207, and T2121001), Key Area Research and Development Program of Guangdong Province, China (Grants No. 2020B0303030001), Beijing Natural Science Foundation (Grant No. Z200009), Innovation Program for Quantum Science and Technology (Grant No. 2021ZD0301800), Strategic Priority Research Program of Chinese Academy of Sciences (Grant No. XDB28000000), and Scientific Instrument Developing Project of Chinese Academy of Sciences (Grant No. YJKYYQ20200041).
	
	
\end{document}